# AN ALTERNATIVE FORM OF HOOGE'S RELATION FOR 1/f NOISE IN SEMICONDUCTOR MATERIALS

FERDINAND GRÜNEIS

*Institute for Applied Stochastic*
*Rudolf von Scholtz Str. 4, 94036 Passau, Germany*
*Email: Ferdinand.Grueneis@t-online.de*

Single quantum dots and other materials exhibit irregular switching between on and off states; these on-off states follow power-law statistics giving rise to 1/f noise. We transfer this phenomenon (also referred to as on-off intermittency) to the generation and recombination (= g-r) process in semiconductor materials. In addition to g-r noise we obtain 1/f noise that can be provided in the form of Hooge's relation. The predicted Hooge coefficient is $\alpha_H = \alpha_X \, \alpha_{im}$ whereby $\alpha_X$ depends on the parameters of the g-r noise and $\alpha_{im}$ on the parameters of the intermittency. Due to the power-law distribution of the on-times, the coefficient $\alpha_{im}$ shows a smooth dependence on time $t$. We also suggest an alternative form of Hooge's 1/f noise formula relating the 1/f noise to the number of centers (such as donor or trap atoms) rather than to the number of charge carriers as defined by Hooge.

## 1. Introduction

1/f noise in semiconductors and other conducting materials continues to attract a great deal of attention; for an overview see the papers by Dutta and Horn [1], Weissman [2], Kogan [3] and Bezrukov, Vandamme and Kish [4]. Despite considerable progress the physical origin of 1/f noise remains controversial. For a homogenous semiconductor, Hooge [5] empirically found

$$\frac{S_{1/f}(f)}{I_0^2} = \frac{\alpha}{N_0} \frac{1}{f} \,. \tag{1}$$

Herein $S_{1/f}(f)$ is the power spectral density of a fluctuating current $I(t)$ with a mean value of $I_0$. $N_0$ is the number of charge carriers in the probe volume and $\alpha$ is the so-called Hooge coefficient. Most researchers agree that 1/f noise in semiconductors and metallic resistors is well described by Eq.(1). However, some doubts concerning the factor $1/N_0$ in (1) have been raised by Weissman [2].

There are several approaches for interpreting 1/f noise in semiconductor materials: McWorther [6] assumes that traps are uniformly distributed through an oxide layer; this leads to a superposition of the g-r spectra summing up to 1/f noise. Kiss and Kleinpenning [7] presented an interpretation of Hooge's 1/f noise formula that relates the intensity of 1/f noise to the Debye screening length. Mihaila [8] discussed microscopic nonlinearities as a possible origin of 1/f noise.

Lykanchikowa found a close relationship between 1/f noise and g-r noise [9]; however, the results could not be reproduced by Hooge [10]. There is an ongoing discussion on whether 1/f noise is caused by number or mobility fluctuations [10-12]. Based on empirical findings, Hooge, Kleinpenning and Vandamme [13] concluded that 1/f noise is a bulk phenomenon; they favor phonon scattering as an origin of 1/f noise. This perspective is also supported by Mihaila [14] who claims that 1/f noise comes from the perpetual equilibrium atomic motion. Musha, Gabor and Minoru [15] investigated the scattering of laser light in quartz and found 1/f fluctuations in the number of phonons.

Recently, 1/f noise has been observed in quantum dots, nanowires and some organic molecules [16-18]. They exhibit periods of bright ("on") and dark ("off") states also denoted by fluorescence intermittency. Most surprising is the observation that these on-off states follow power-law statistics; blinking occurs over a wide range of timescales from µs to minutes. An interpretation for the phenomenon of intermittency in quantum dots and other materials remains lacking.

This paper transfers these findings to semiconductor materials in applying intermittency to the (non-radiative) g-r process. This leads to an alternative form of Hooge's 1/f noise formula relating 1/f noise to the number of centers rather than to the number of charge carriers as defined in (1). The 1/f noise exhibits a smooth dependence on the time.

The possible mechanisms relating 1/f noise to an intermittent g-r process were already investigated in a previous paper [19]; however the results derived apply only in the special cases of low and high ionization and only in the cases of small intermissions. This paper presents a general formulation for the spectral features of the intermittent g-r process.

$N_0$ conduction electrons
$N_D$ fully ionized donor atoms

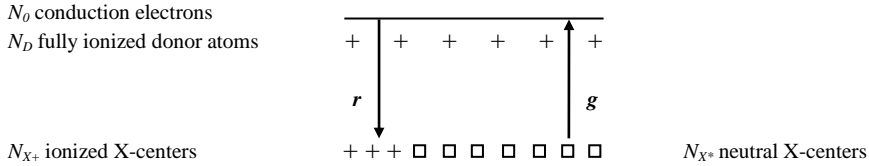

$N_{X_+}$ ionized X-centers                              $N_{X^*}$ neutral X-centers

Fig. 1. The energy band system and electron transitions involved in the fluctuations of conduction electrons in a two-level system. $N_X$ is the number of X-centers, $N_D$ is the number of fully ionized shallow donor atoms. Under steady state conditions, $N_{X_+}$ is the number of ionized X-centers and $N_{X^*}$ is the number of neutral X-centers. $N_0$ is the number of conduction electrons.

## 2. Fluctuations in a Semiconductor

This paper's considerations are confined to g-r noise in a two-level system of a doped n-type semiconductor. Thermal noise that is always present will be neglected. For steady state conditions the generation and recombination rate are denoted by $g_0$ and $r_0$ respectively; $N_0$ is the steady state value for conduction electrons. The number fluctuations of conduction electrons tend to be Gaussian distributed with $\overline{\Delta N^2}$ being the mean square fluctuations of conduction electrons about $N_0$. Under steady state conditions, the master equation approach leads to the following expressions [20-23]

$$g_0 = r_0 = \overline{\Delta N^2} / \tau_{gr} \tag{2}$$

$$S_{gr}(f) = \left(\frac{I_0}{N_0}\right)^2 \frac{4 g_0 \tau_{gr}^2}{1+(2\pi f \tau_{gr})^2}. \tag{3}$$

Herein, $\tau_{gr}$ is the relaxation time of the g-r process and $S_{gr}(f)$ is the power spectral density of the current fluctuations due to an applied current $I_0$.

### 2.1. *Fluctuations in a two-level system*

This report considers only transitions between the level of X-centers and the conduction bands as producing fluctuations (see Fig. 1). For example, X-centers may be donor or trap atoms. Let the semiconductor volume be 1cm³ so that the number of electrons, donors, traps and holes coincides with their concentration. The number of X-centers is denoted by $N_X$ and the number of fully ionized shallow donor atoms is denoted by $N_D$. The mean number of ionized X-centers is $N_{X+}$, of neutral X-centers is $N_{X*} = N_X - N_{X+}$ and of conduction electrons is $N_0 = N_{X+} + N_D$. Under steady state conditions, the rate of generation and recombination is

$$g_0 = \gamma N_{X*} \tag{4}$$

and

$$r_0 = \rho N_0 N_{X+}. \tag{5}$$

Herein, $\gamma$ and $\rho$ are the coefficients of generation and recombination respectively. Defining the generation and recombination rate by $1/\tau_g$ and $1/\tau_r$ respectively, these coefficients can be expressed as

$$1/\tau_g = \gamma \tag{6}$$

and

$$1/\tau_r = \rho N_0 + \rho N_{X+}. \tag{7}$$

We define the two contributions on the r.h.s. separately by

$$1/\tau_{r+} = \rho N_0 \tag{8}$$

and

$$1/\tau_{r_c} = \rho N_{X+} \tag{9}$$

leading to

$$1/\tau_r = 1/\tau_{r+} + 1/\tau_{r_c}. \tag{10}$$

Herein, $1/\tau_{r+}$ and $1/\tau_{r_c}$ can be interpreted as the mean rate for the recombination due to the ionized X-centers and due to conduction electrons respectively. Using these definitions, the g-r relaxation time is

$$1/\tau_{gr} = 1/\tau_g + 1/\tau_{r+} + 1/\tau_{r_c}. \tag{11}$$

By virtue of Eq. (2), the mean square fluctuations of conduction electrons is determined by

$$\frac{1}{\overline{\Delta N^2}} = \frac{1}{N_{X*}} + \frac{1}{N_{X+}} + \frac{1}{N_0}. \tag{12}$$

Applying Eqs. (6) and (8), Eqs.(4) and (5) are transformed to

$$g_0 = N_{X*}/\tau_g \tag{13}$$

and

$$r_0 = N_{X+}/\tau_{r+}. \tag{14}$$

Under steady state conditions $g_0 = r_0$ leading to

$$\frac{\tau_g}{\tau_{r+}} = \frac{N_X - N_{X+}}{N_{X+}}. \tag{15}$$

Using this, the fraction of ionized X-centers can be expressed by

$$f_{X+} = \frac{N_{X+}}{N_X} = \frac{\tau_{r+}}{\tau_X} \tag{16}$$

whereby

$$\tau_X = \tau_g + \tau_{r+}. \tag{17}$$

Combining Eqs. (14) and (16), under steady state conditions

$$g_0 = r_0 = \frac{N_X}{\tau_X}. \tag{18}$$

For further use we note an alternative form of the g-r noise: applying Eqs. (2) and (18), Eq. (3) can be provided as

$$\frac{S_{gr}(f)}{I_0^2} = \frac{\tau_X}{N_X}\left(\frac{\overline{\Delta N^2}}{N_0}\right)^2 \frac{4}{1+(2\pi f \tau_{gr})^2} \qquad (19)$$

relating the g-r noise to the number of X-centers.

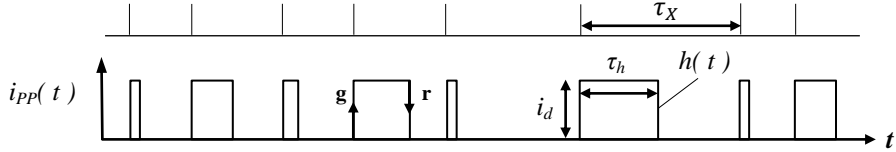

Fig. 2. Top: The spike train indicates the time points for the generation of an electron at a single X-center; the time between the successive generations is $\tau_X$. Alternatively, $\tau_X$ can be interpreted as the time between successive recombinations. Bottom: Current fluctuations due to a single X-center. The time series $i_{PP}(t)$ consists of successive rectangular current pulses $h(t)$ with lifetime $\tau_h$ and amplitude $i_d$. The pulse $h(t)$ represents an elementary g-r process: after generation at a single X-center, an electron remains for the lifetime $\tau_h$ in the conduction band before it recombines to an arbitrary ionized X-center. A mean drift current is $i_d = e\mu E_0/L$ where $e$ is the elementary charge, $\mu$ is the mobility, $E_0$ is an applied electric field and $L$ is the length of the sample.

## 2.2. *The noise contribution due to a single X-center*

To compare with <u>single</u> quantum dots and other material, the noise contribution due to a <u>single</u> X-center is considered; this is defined by

$$S_X(f) \equiv \frac{S_{gr}(f)}{N_X}. \qquad (20)$$

We show that this noise contribution can be interpreted as a random succession of elementary g-r pulses. According to Eq. (18), the generation rate due to a single X-center is $g_0/N_X = 1/\tau_X$ where $\tau_X$ is the time between successive generations[1] (see the top of Fig. 2). As shown on the bottom of Fig. 2 each spike triggers an elementary g-r pulse $h(t)$ leading to a time series $i_{PP}(t)$. This random succession of g-r pulses can be interpreted as shot noise. Applying Carson's theorem for elementary events [24] the power spectral density of this pulse train is obtained using

$$S_X(f) = \frac{2}{\tau_X}\overline{|H(f)|^2}. \qquad (21)$$

Herein, $H(f)$ is the Fourier transform of $h(t)$. As is well-known [21-23], the lifetime $\tau_h$ of conduction electrons is exponentially distributed yielding

$$\overline{|H(f)|^2} = 2\left|\overline{H(f)}\right|^2 = \frac{2 i_d^2 \tau_{gr}^2}{1+(2\pi f \tau_{gr})^2}. \qquad (22)$$

According to Eq. (20) the noise contribution due to the $N_X$ centers is

$$S_{gr}(f) = N_X S_X(f) = 4\frac{N_X}{\tau_X}\frac{i_d^2 \tau_{gr}^2}{1+(2\pi f \tau_{gr})^2}. \qquad (23)$$

Taking into account Eq. (18) and considering that a current $I_0 = N_0\, i_d$ Eq. (23) is in accordance with Eq. (3). Hence, under steady state conditions, the shot noise interpretation of the g-r noise is consistent with the master equation approach to the g-r noise.

---

[1] To avoid extended mathematical formula, we do not distinguish between a statistical variable and its mean value.

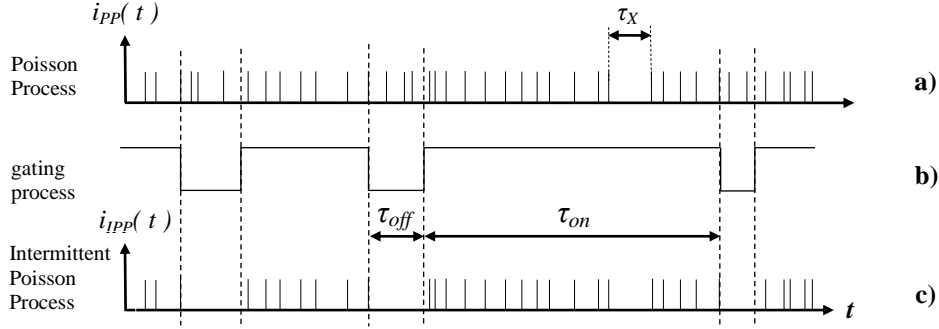

Fig. 3.a. Schematic plot of a Poisson Process (= PP) with time series $i_{PP}(t)$ representing the occurrence of successive elementary g-r pulses due to a single X-center; $\tau_X$ is the time between successive g-r pulses. Each spike triggers an elementary g-r current pulse $h(t)$ as is seen in Fig. 2 (not indicated in this Figure). Fig. 3.b. The Poisson Process is gated by a two-state process with states $\tau_{off}$ (= intermission) and $\tau_{on}$ (= duration of a cluster). Fig. 3.c. The Intermittent Poisson Process (= IPP) is characterized by intermissions followed by fluctuating clusters. Each spike triggers an elementary g-r pulse $h(t)$ (not indicated in this Figure) leading to the time series $i_{IPP}(t)$.

## 3. An intermittent g-r process as a possible origin of 1/f noise

As mentioned in the Introduction, blinking of single quantum dots and other materials is characterized by "on-off" states that are both power-law distributed such as $1/t^{\mu_{on/off}}$ with an exponent $\mu_{on/off}$ ranging from 1.2 to 2.0; correspondingly, their spectra exhibit two scaling regions with two different slopes [16-18]. We transfer this phenomenon to semiconductor materials in assuming that the g-r process is also controlled by such "on-off" states. However, as a rule, 1/f noise in semiconductors exhibits only one scaling region; this can be explained only if the on-state is power-law distributed whereas the off-state is exponentially distributed (Case Ia in [18]). By analogy to a blinking <u>single</u> quantum dot, we regard a <u>single</u> X-center. The g-r process of such a single X-center is assumed to be intermitted by a gating function with $\tau_{off}$ being the off-time (see Fig. 3.b). In this way the Intermittent Poisson Process (= IPP) is obtained as shown in Fig. 3.c.

As in quantum dots and other materials, the on-times are assumed to be power-law distribution such as $1/t^{\mu_{on}}$. This leads to a finite and random number of spikes in so-called clusters (see Fig. 3.c); these number fluctuations can be described by a cluster size distribution $q_n$ that follows the power-law distribution of on-times

$$q_n \propto 1/n^{\mu_{on}}. \tag{24}$$

Herein $n = 1, 2,\dots M_t$ where $M_t$ is the maximum number of spikes in a cluster; hence $q_n$ is a truncated Zeta or Zipf distribution [25]. The exponent $\mu_{on}$ will be specified in the next section and $M_t$ in Section 3.4. Denoting the mean number of spikes in a cluster by $\overline{N_c} = \sum n\, q_n$ the mean on-time (= the mean duration of a cluster) is obtained by

$$\tau_{on} = \overline{N_c}\tau_X. \tag{25}$$

### 3.1. *1/f noise and g-r noise due to a single X-center*

The Intermittent Poisson Processes $i_{IPP}(t)$ in Fig. 3.c represents the current fluctuations due to a single X-center; each spike triggers an elementary g-r current pulse $h(t)$ with Fourier transform $H(f)$. The power spectral density of current fluctuations is given by [18, 26]

$$S_{IPP}(f) = \frac{2}{\tau_{on} + \tau_{off}} \left\{ \overline{N_c} \; \overline{|H(f)|^2} \; + \left| \overline{H(f)} \right|^2 \Phi_X(f) \right\}. \tag{26.a}$$

Using Eqs. (21), (22) and (25) this can be rewritten as

$$S_{IPP}(f) = S_X(f) \left\{ \frac{\overline{N_c}}{\overline{N_c} + \frac{\tau_{off}}{\tau_X}} + \frac{1}{2} \frac{\Phi_X(f)}{\overline{N_c} + \frac{\tau_{off}}{\tau_X}} \right\}. \tag{26.b}$$

The second term in curly brackets is 1/f noise with spectral function (see also Appendix)

$$\Phi_X(f) \approx \left( \frac{\tau_{off}}{\tau_X} \right)^2 \frac{C_{im}}{(f \tau_X)^b} \tag{27}$$

applying for $\tau_{off} \ll \tau_{on}$; our considerations in Section 3.4 suggest that this condition is met for most materials. 1/f noise scales within the lower and upper cut-off frequency

$$f_l \approx 1/2 M_t \tau_X \tag{28.a}$$

and

$$f_u \approx 1/2\pi\tau_X. \tag{28.b}$$

Beyond $f_u$ the 1/f shape rapidly approaches zero; below $f_l$ the 1/f noise reaches a plateau. Hence, the scaling region of the 1/f noise is

$$f_u / f_l \approx M_t. \tag{29}$$

The numerical constant $C_{im}$ and slope $b$ in Eq. (27) and the exponent $\mu_{on}$ in Eq. (24) depend on the scaling region $M_t$ and on the quotient $\tau_{off}/\tau_X$. A discussion of the behavior of the numerical constant $C_{im}$ is beyond the scope of this paper; our considerations are confined to a well extended scaling region and to a slope $b \approx 1$: under these conditions we find $C_{im} \approx 0.01$.

The procedure for obtaining slope $b$ and exponent $\mu_{on}$ as a function of normalized intermission $\tau_{off}/\tau_X$ and of the scaling region $M_t$ is described in the Appendix. Table 1 summarizes the results for a scaling region $M_t = 10^{15}$.

Table 1. The slope $b$ and the exponent $\mu_{on}$ for several values of $\tau_{off}/\tau_X$. The scaling region $M_t = 10^{15}$.

| $\tau_{off}/\tau_X$ | $10^{-1}$ | $10^{0}$ | $10^{1}$ | $10^{2}$ | $10^{3}$ | $10^{4}$ |
|---|---|---|---|---|---|---|
| $\mu_{on}$ | 2.00 | 2.00 | 1.96 | 1.90 | 1.82 | 1.74 |
| $b$ | 0.81 | 0.81 | 0.92 | 1.02 | 1.10 | 1.18 |
| $\overline{N_c}$ | 25 | 25 | 44 | 168 | 1385 | 10250 |
| $\tau_{off}/\tau_{on}$ | 0.04 | 0.04 | 0.23 | 0.60 | 0.73 | 0.98 |

Similar to blinking of single quantum dots and other materials the exponent $\mu_{on} \leq 2$. For $\mu_{on} = 2$, the mean number of spikes in a cluster is obtained by

$$\overline{N_c} \approx \frac{6}{\pi^2} (\ln M_t + C_E) \tag{30}$$

where $C_E \approx 0.5772\ldots$ is Euler's constant. For $\mu_{on} < 2$ the sum $\overline{N_c} = \sum n \, q_n$ is replaced by integration leading to

$$\overline{N_c} \approx \frac{6}{\pi^2} \frac{1 - \mu_{on}}{2 - \mu_{on}} \frac{M_t^{2 - \mu_{on}} - 1}{M_t^{1 - \mu_{on}} - 1}. \tag{31}$$

The factor $\frac{6}{\pi^2}$ guarantees the equivalence to Eq. (30) for $\mu_{on} \to 2$. Using Eqs. (30) and (31), Table 1 shows the mean cluster size $\overline{N_c}$ for several values of $\mu_{on}$ and for a scaling region $M_t = 10^{15}$. Table 1 also shows $\tau_{off}/\tau_{on} = \tau_{off}/\overline{N_c}\tau_X$.

### 3.2. *1/f noise and g-r noise in the probe volume*

Under the supposition that the noise contributions due to the $N_X$ single X-centers are statistically independent, the total noise in the probe volume is obtained by

$$S_{tot}(f) = N_X S_{IPP}(f). \tag{32}$$

Substituting herein Eq. (26.b) obtains

$$S_{tot}(f) = \beta_{im} S_{gr}(f) + S_{1/f}(f). \tag{33}$$

The first term on the r.h.s. is reduced g-r noise containing a pre-factor

$$\beta_{im} = \overline{N_c} / \left( \overline{N_c} + \frac{\tau_{off}}{\tau_X} \right). \tag{34}$$

This term is investigated below. The second term is 1/f noise

$$S_{1/f}(f) = \tfrac{1}{2} S_{gr}(f) \Phi_X(f) / \left( \overline{N_c} + \frac{\tau_{off}}{\tau_X} \right). \tag{35}$$

Using Eqs. (19) and (27), the 1/f noise can be given a generalized form of Hooge's relation

$$\frac{S_{1/f}(f)}{I_0{}^2} = \frac{\alpha_H}{N_0} \frac{\tau_X}{(f\tau_X)^b}. \tag{36}$$

For $b \to 1$, this reduces to

$$\frac{S_{1/f}(f)}{I_0{}^2} = \frac{\alpha_H}{N_0} \frac{1}{f} \tag{37}$$

attaining the form of Hooge's relation in Eq. (1). The predicted Hooge coefficient is

$$\alpha_H = \alpha_{im} \alpha_X. \tag{38}$$

Herein, the coefficient

$$\alpha_{im} = 2C_{im} \left( \frac{\tau_{off}}{\tau_X} \right)^2 / \left( \overline{N_c} + \frac{\tau_{off}}{\tau_X} \right) \tag{39}$$

depends on the parameters of intermission as well as on cluster formation; $\alpha_{im}$ is investigated below. The coefficient

$$\alpha_X = \frac{\overline{\Delta N^2}}{N_X} \frac{\overline{\Delta N^2}}{N_0} \tag{40}$$

is determined by the parameters of the g-r process.

### 3.3. *An alternative form of Hooge's relation for 1/f noise*

According to Eq. (40)

$$\frac{\alpha_X}{N_0} = \frac{1}{N_X} \left( \frac{\overline{\Delta N^2}}{N_0} \right)^2 \tag{41}$$

transforming Eq. (37) to

$$\frac{S_{1/f}(f)}{I_0{}^2} = \frac{\gamma_H}{N_X} \frac{1}{f}. \tag{42}$$

Herein

$$\gamma_H = \alpha_{im} \left(\frac{\overline{\Delta N^2}}{N_0}\right)^2 \tag{43}$$

is the alternative Hooge coefficient relating 1/f noise to $N_X$ (= the number of X-centers) rather than to $N_0$ (= the number of charge carriers) as is defined by Hooge in Eq. (1). By analogy to Eq. (38) we express Eq. (43) in the form

$$\gamma_H = \alpha_{im} \gamma_X \tag{44}$$

leading to

$$\gamma_X = \left(\frac{\overline{\Delta N^2}}{N_0}\right)^2. \tag{45}$$

Using Eq. (41), the coefficient $\alpha_X$ can be written as

$$\alpha_X = \frac{N_0}{N_X} \left(\frac{\overline{\Delta N^2}}{N_0}\right)^2. \tag{46}$$

Considering that $N_0 = N_{X+} + N_D$ the first term on the r.h.s. is

$$\frac{N_0}{N_X} = f_{X+} + R_X \tag{47}$$

where $f_{X+} = N_{X+} / N_X$ is the ionization factor of the X-centers and $R_X = N_D / N_X$ is the normalized number of shallow donors. Applying Eq. (12) we obtain the relative fluctuations of charge carriers by

$$\frac{\overline{\Delta N^2}}{N_0} = \frac{f_{X+}(1-f_{X+})}{f_{X+}(2-f_{X+})+R_X}. \tag{48}$$

Substituting Eqs. (47-48) into Eqs. (45-46) results in

$$\alpha_X = (f_{X+} + R_X) \left(\frac{f_{X+}(1-f_{X+})}{f_{X+}(2-f_{X+})+R_X}\right)^2. \tag{49}$$

and

$$\gamma_X = \left(\frac{f_{X+}(1-f_{X+})}{f_{X+}(2-f_{X+})+R_X}\right)^2. \tag{50}$$

Fig. 4 shows the coefficients $\alpha_X$ and $\gamma_X$ as a function of $f_{X+}$ for several values of $R_X$.

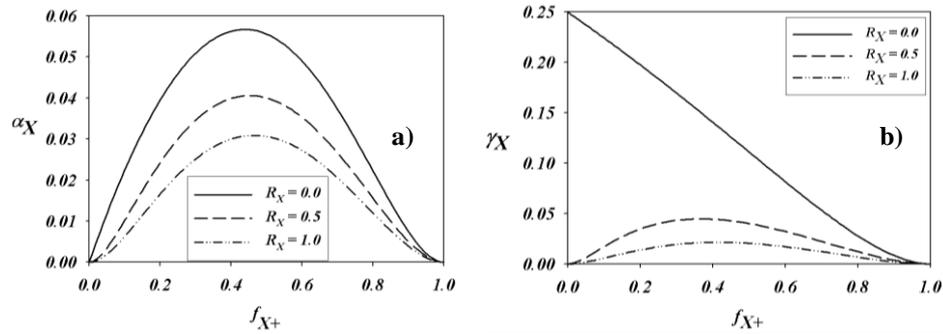

Fig. 4.a and b. The coefficients $\alpha_X$ and $\gamma_X$ as a function of $f_{X+}$ for several values of $R_X$.

The magnitude of 1/f noise may also be represented by a coefficient

$$C_X = \frac{\gamma_X}{N_X} = \frac{\alpha_X}{N_0} = \left(\frac{\overline{\Delta N^2}}{N_0}\right)^2 \Big/ N_X \tag{51}$$

leading to

$$\frac{S_{1/f}(f)}{I_0^2} = C_X \frac{\alpha_{im}}{f}. \tag{52}$$

Table 2 summarizes approximations of $C_X$ for low/high ionization and for low/high doping with additional shallow donors.

Table 2. Approximations of $C_X$ for low/high ionization and low/high doping with additional shallow donors.

|  | $N_D = 0$ | $N_D \gg N_X$ |
|---|---|---|
| $f_{X+} \to 0$ | $C_X \approx \frac{1}{4N_X}$ | $C_X \approx \frac{1}{N_X}\left(\frac{N_{X+}}{N_D}\right)^2$ |
| $f_{X+} \to 1$ | $C_X \approx \frac{(1-f_{X+})^2}{4N_X}$ | $C_X \approx \frac{1}{N_X}\left(\frac{N_{X+}}{N_D}\right)^2$ |

### 3.4.  *Dependence of 1/f noise and g-r noise on time*

What remains to be considered is the behavior of the coefficients $\alpha_{im}$ and $\beta_{im}$ that is the pre-factors to 1/f noise and g-r noise (see Eqs. (34) and (39)).

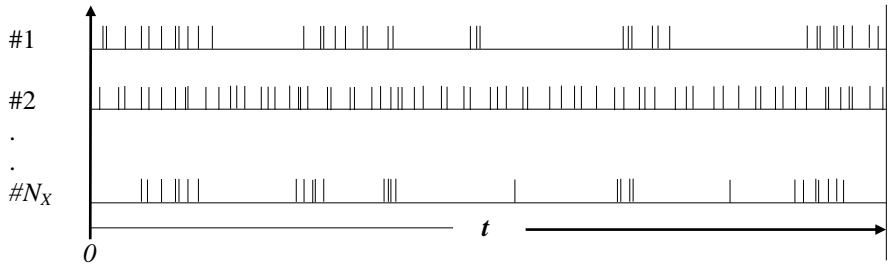

Fig. 5. The $N_X$ centers starting simultaneously at $t = 0$ with the generation of the IPP; each spike triggers an elementary g-r current pulse $h(t)$ as is seen in Fig. 2 (not indicated in this Figure). The time between spikes in a cluster is $\tau_X$. The X-centers are numbered #1, #2, ....#$N_X$. The X-center #2 shows a cluster without any intermissions during time $t$.

We define $t = 0$ as the time when the $N_X$ centers simultaneously start generating an Intermittent Poisson Process (= IPP); for an illustration see Fig. 5. Consequently, the time $t$ can be interpreted as the age of the semiconductor material; at $t = 0$ the material is assumed to be in thermal equilibrium.

Among these $N_X$ centers there will be a certain number of clusters without any intermissions during $t$; as an example see X-center #2 in Fig. 5. In such clusters, the mean number of events occurring within $t$ is denoted by $M_t$; this obeys Poisson statistics leading to

$$M_t = t/\tau_X. \tag{53}$$

The spread about $M_t$ is described by the normalized standard deviation

$$\frac{\sigma_{M_t}}{M_t} = \sqrt{\frac{\tau_X}{t}} \tag{54}$$

approaching zero for large $t$. According to (24), the probability for clusters without any intermissions is $q_{M_t} \ll 1$. For this reason, the spread about $M_t$ is much smaller than Eq. (54), which was assured by simulations of the IPP [26].

Consequently, all of the quantities containing $M_t$ depend on time; this applies to the lower limit of 1/f noise in Eq. (28.a), the scaling region in Eq. (29) and the mean cluster size in Eq. (31). Using Eqs. (31) and (53), a mean cluster size as a function of time is obtained by

$$\overline{N_c(t)} \approx \frac{6}{\pi^2} \frac{1-\mu_{on}}{2-\mu_{on}} \frac{\left(\frac{t}{\tau_X}\right)^{2-\mu_{on}}-1}{\left(\frac{t}{\tau_X}\right)^{1-\mu_{on}}-1}. \tag{55}$$

Hence, the pre-factors of 1/f noise and g-r noise $\alpha_{im}$ and $\beta_{im}$ also depend on time. Substituting Eq. (55) into Eqs. (34) and (39) yields

$$\alpha_{im}(t) = 2C_{im}\left(\frac{\tau_{off}}{\tau_X}\right)^2 / \left\{\overline{N_c(t)} + \frac{\tau_{off}}{\tau_X}\right\} \tag{56.a}$$

and

$$\beta_{im}(t) = \overline{N_c(t)} / \left\{\overline{N_c(t)} + \frac{\tau_{off}}{\tau_X}\right\}. \tag{56.b}$$

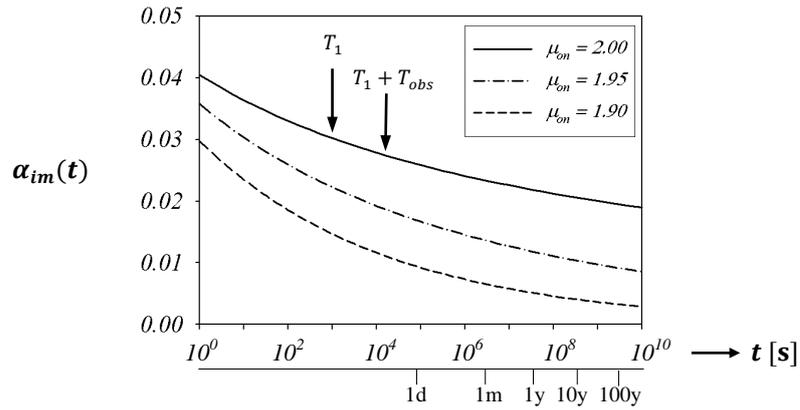

Fig. 6. The coefficient $\alpha_{im}(t)$ for $\tau_X = 10^{-5}$s, $\tau_{off}/\tau_X = 5$, and several values of $\mu_{on}$. Also indicated is the beginning of an observation at $T_1$ lasting for observation time $T_{obs}$. Herein, d = day, m = month and y = year.

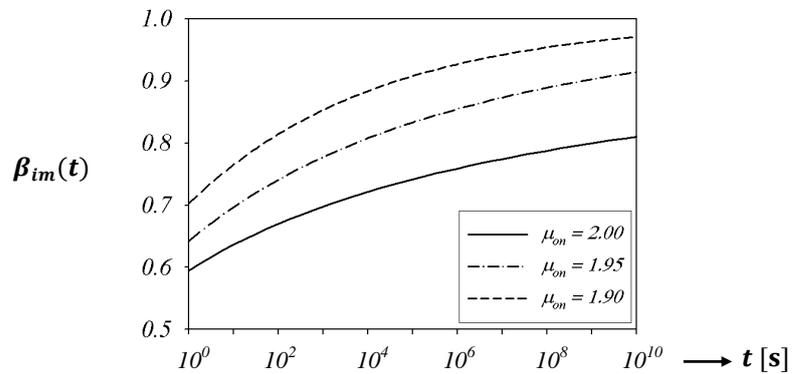

Fig. 7. The coefficient $\beta_{im}(t)$ for $\tau_X = 10^{-5}$s and $\tau_{off}/\tau_X = 5$ and for several values of $\mu_{on}$.

These coefficients contain the quotient $\tau_{off}/\tau_X$ as a free parameter. This quotient is adjusted in such a way that the maximum value of the predicted Hooge coefficient $\alpha_H = \alpha_{im}\alpha_X$ is equal to the maximum empirical value in semiconductors of approximately $2\cdot10^{-3}$ [13]. This leads to $\tau_{off}/\tau_X \leq 5$ suggesting that intermissions suppress only a few generations. Recently, the Hooge coefficient has been found to be considerably higher in carbon nanotubes: the highest value $\alpha \approx 0.2$ estimated by [27] could not be reproduced by [28] and has been corrected to $\alpha \approx 0.02$ leading to $\tau_{off}/\tau_X \leq 8$.

Choosing $\tau_X = 10^{-5}$s and $\tau_{off}/\tau_X = 5$, Figs. 6 and 7 show $\alpha_{im}(t)$ and $\beta_{im}(t)$ for several values of $\mu_{on}$. 1/f noise decreases and reduced g-r noise increases with time $t$. Eventually, for very long times, 1/f noise is negligibly small whereas reduced g-r noise approaches its maximum value provided by Eq. (3).

In addition, also slope $b$ and exponent $\mu_{on}$ depend on time. Using Eqs. (53), (A.13) and (A.14), slope $b$ as a function of time can be written as (see Appendix)

$$\tau_{off} \leq \tau_X: \qquad b \approx 1 - \frac{\log\frac{6}{\pi^2}\left[\ln\left(\frac{t}{\tau_X}\right)+C_E\right]^2}{\log\left(\frac{t}{\tau_X}\right)} \qquad (57.a)$$

$$\tau_X < \tau_{off} \ll \tau_{on}: \qquad b \approx 1 - \frac{\log\frac{6}{\pi^2}\left[\ln\left(\frac{t}{\tau_X}\right)+C_E\right]^2}{\log\left(\frac{t}{\tau_X}\right)} + \frac{2\log\left(\frac{\tau_{off}}{\tau_X}\right)}{\log\left(\frac{t}{\tau_X}\right)} \qquad (57.b)$$

showing a rather slow convergence to $b = 1$ from below. Under these conditions, the exponent $\mu_{on} = 2$ independent on time. However, for $\tau_{off} \leq \tau_{on}$, Eqs. (A.15) and (A.16) lead to

$$b(t) \approx 1 + \frac{\log\left(\frac{\tau_{off}}{\tau_X}\right)}{\log\left(\frac{t}{\tau_X}\right)} \qquad (58.a)$$

and

$$\mu_{on}(t) \approx 2 - \frac{\log\left(\frac{\tau_{off}}{\tau_X}\right)}{\log\left(\frac{t}{\tau_X}\right)}. \qquad (58.b)$$

For long times, slope $b(t)$ approaches $b = 1$ from above and the exponent $\mu_{on}(t)$ converges to $\mu_{on} = 2$.

### 3.5. *Possible pitfalls when measuring the Hooge coefficient*

Due to the component $\alpha_{im}(t)$ in Eq. (56.a) the Hooge coefficient in Eq. (37) also depends on time. Let $T_1$ be the time at the beginning of an observation and $T_{obs}$ the time of observation. During the time of observation the coefficient $\alpha_{im}$ decreases; hence, we are measuring an average of $\alpha_{im}(t)$ over the time of observation

$$< \alpha_{im} > = \frac{1}{T_{obs}} \int_{T_1}^{T_1+T_{obs}} \alpha_{im}(t)\, dt. \qquad (59)$$

Obviously, for $T_1 \ll T_{obs}$ (say $T_1 = 10^3 s$ and $T_{obs} = 10^4 s$ as is seen in Fig. 6) averages of $\alpha_{im}$ for the same material may be quite different if the measurements start at different times and if the time window of the observation is prolonged or shortened. However, for $T_1 \gg T_{obs}$ (say $T_1 = 10$ years and $T_{obs} = 10^4 s$) the small dependence of $\alpha_{im}$ on time may lead to a misinterpretation of $\alpha_{im}$ as a constant independent on time.

Even more complex is the situation if we compare 1/f noise in different types of semiconductor materials: now – in addition to the time dependent component $\alpha_{im}$ – also the coefficient $\alpha_X / N_0$ comes into play. According to (49), the ionization of X-centers (controlled by the temperature), the degree of doping with additional shallow donors and the number of X-centers of the different semiconductor materials must be considered.

## 4. Results and Discussions

The present paper provides possible mechanisms relating $1/f^b$ noise to an intermittent g-r process. Our approach is inspired by recent measurement of blinking single quantum dots, nanowires and some organic molecules; they exhibit fluorescence intermittency switching irregularly between the on and off states which are power-law distributed.

We transfer these findings to semiconductor materials in assuming that the g-r process of a single X-center (like a trap or donor atom) is also gated by such on-off states. In this way, an intermittent g-r process is introduced where intermissions (the off-states) are followed by a fluctuating number of g-r pulses representing the on-state. This paper demonstrates that a slope $0.8 \leq b \leq 1.2$ originates from on-states that are power-law distributed such as $1/t^{\mu_{on}}$ with $1.7 \leq \mu_{on} \leq 2$.

Hooge's relation in Eq. (1) applies for a pure $1/f$ shape. However, a slope $b$ slightly different from 1 is usually observed [8, 29]. Our derivations yielding a $1/f^b$ shape include such deviations leading to a generalized form of Hooge's relation. For $b \rightarrow 1$, a comparison with Eq. (1) leads to the predicted Hooge coefficient $\alpha_H = \alpha_X \alpha_{im}$; herein

- the coefficient $\alpha_X$ comprises the impact of g-r noise on $1/f$ noise; it depends on the ionization factor of X-centers and on doping with additional shallow donors. Such a close relationship between the $1/f$ noise and the g-r noise was already described by Lukyanchikowa [9] and Van Vliet [30] and is confirmed by the present model.
- the coefficient $\alpha_{im}$ describes the impact of intermission and cluster formation. Due to the power-law distribution of on-times $\alpha_{im}$ shows a smooth dependence on time.

Some doubts concerning the factor $1/N_0$ in Hooge's relation in Eq. (1) have been raised by Weissman [2]. Hooge's relation suggests that some independent fluctuations are occurring on each of the mobile carriers – this is the interpretation of the Hooge group [13]. Weissman however argues that any fluctuations tied to individual mobile carriers cannot persist for times longer than the carrier remains in the sample. This seemingly paradoxical situation is circumvented by the introduction of an intermittent g-r process. Long times caused by "long" clusters can be sensed though the charge carriers causing these clusters recombined long time ago. Moreover, the factor $1/N_0$ is not compelling for a description of the $1/f$ noise in semiconductor materials: we suggest an alternative form of Hooge's formula relating $1/f$ noise to $N_X$ (= number of X-centers) rather than to $N_0$ (= number of charge carriers) as defined by Hooge in Eq. (1).

Due to the power-law distribution of on-times the parameters determining $1/f^b$ noise and g-r noise show a smooth dependence on time $t$. When measuring the Hooge coefficient $\alpha_H$ we obtain an average of $\alpha_H$ over the time of observation. This may explain the spread of Hooge coefficients observed for the same and even more for different semiconductor materials [12, 31]. We also investigate the slope $b$ as a function of time approaching $b = 1$ for long times.

Similar to quantum dots and other materials, the physical origin of intermissions in a semiconductor material remains open to question. However, as is well-known, the g-r process in a semiconductor is phonon-induced; a theory describing this mechanism has been established by Shockley [32]. Hence, the intermittent g-r process generating $1/f$ noise is also phonon-induced. Indeed, there is ample empirical evidence that electron-phonon scattering is the dominate mechanism generating $1/f$ noise in semiconductors [13-15]. As was already mentioned in a previous paper [19], this suggests that the intermissions (= off-states) originate from phonons due to an unknown mechanism. It is possible that non-linear interactions between different phonon modes may cause a short break down of a mode for example by interference. Such behavior is supported by Mihaila [8] who considered nonlinearity induced energy exchange between eigen-modes as essential to distribute thermal energy between different modes. It is also supported by Ruseckas and Kaulakys [33] who showed that differential equations with non-linear terms may cause intermittency accompanied by 1/f fluctuations.

This research did not receive any specific grant from funding agencies in the public, commercial, or not-for-profit sectors.

## Appendix. The Behavior of Slope *b* for Large Scaling Regions

Introducing the ratio

$$r = \frac{\tau_{off}}{\tau_X} \tag{A.1}$$

the $1/f^b$ noise in Eq. (27) can be expressed by

$$\Phi_X(f) \approx r^2 \frac{c_{im}}{(f\tau_X)^b}. \tag{A.2}$$

This relation applies to $\tau_{off} \ll \tau_{on}$ (equivalent to $r \ll \overline{N_c}$). However, for $\tau_{off} \leq \tau_{on}$ (equivalent to $r \leq \overline{N_c}$) this equation has to be replaced by

$$\Phi_X(f) \approx \frac{C_{1/f}}{(f\tau_X)^b} \tag{A.3}$$

where $C_{1/f} = 0.3$ is a numerical constant [34]. The considerations in Section 3.4 suggest that Eq. (A.2) applies for most semiconductor materials. Slope *b* is determined by (see Fig. 8)

$$b = -\frac{\log \Phi_X(f_l) - \log \Phi_X(f_u)}{\log f_l - \log f_u}. \tag{A.4}$$

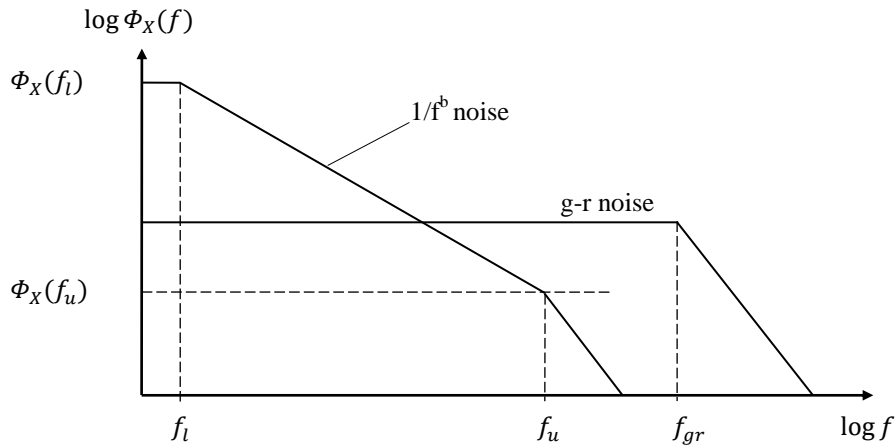

Fig. 8. Schematic plot of $1/f^b$ noise component $\Phi_X(f)$. $f_l$ and $f_u$ is the lower and upper cut-off frequency of $1/f^b$ noise respectively. The upper frequency limit of g-r noise $f_{gr} = 1/2\pi\tau_{gr}$ is well above $f_u$.

The magnitude of the 1/f noise at $f_l$ can be expressed by [26]

$$\Phi_X(f_l) \approx \Phi_X(0) \approx \overline{N_c^2} \left(\frac{r}{r+\overline{N_c}}\right)^2. \tag{A.5}$$

Herein $\overline{N_c^2} = \sum n^2 q_n$ is the second moment of cluster size. Replacing the sum by integration we obtain for $\mu_{on} = 2$

$$\overline{N_c^2} \approx \frac{6}{\pi^2} M_t \tag{A.6}$$

and by analogy to Eq. (31) for $\mu_{on} < 2$

$$\overline{N_c^2} \approx \frac{6}{\pi^2} \frac{1-\mu_{on}}{3-\mu_{on}} \frac{M_t^{3-\mu_{on}}-1}{M_t^{1-\mu_{on}}-1}. \tag{A.7}$$

The magnitude of the 1/f noise at $f_u$ is for

$r \leq 1$: $\qquad\qquad \Phi_X(f_u) \approx 2r^2(q_1 - q_2 - q_1^2) \tag{A.8}$

and for

$r > 1$: $\qquad\qquad \Phi_X(f_u) \approx 2(1 - q_1 - \frac{1}{r}). \tag{A.9}$

Using Eq. (29), we distinguish two cases derived from Eq. (A.4) which apply to

$r \leq 1$: $\qquad\qquad b \approx \log\left\{\frac{\overline{N_c^2}}{\overline{N_c}^2}\right\} / \log M_t \tag{A.10}$

and to

$1 < r$: $\qquad\qquad b \approx \log\left\{\overline{N_c^2} \left(\frac{r}{r+\overline{N_c}}\right)^2\right\} / \log M_t. \tag{A.11}$

The latter relation includes a subcase which applies to

$1 < r \ll \overline{N_c}$: $\qquad\qquad b \approx \log\left\{r^2 \frac{\overline{N_c^2}}{\overline{N_c}^2}\right\} / \log M_t. \tag{A.12}$

Based on Eq. (A.11), Fig. 9 shows the slope $b$ as a function of the exponent $\mu_{on}$ for several values of $r$ and for a scaling region $M_t = 10^{15}$.

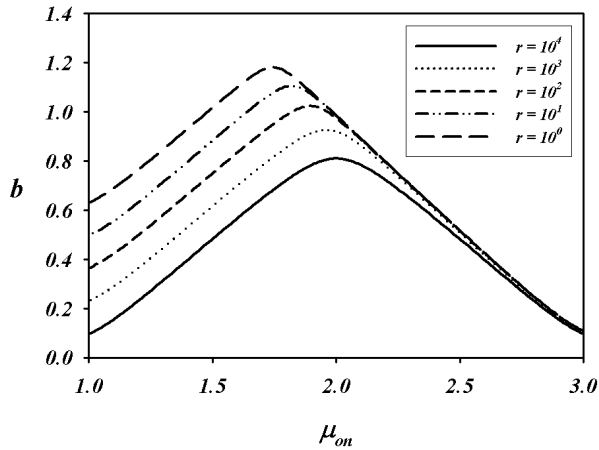

Fig. 9. The slope $b$ according to Eq. (A.11) as a function of the exponent $\mu_{on}$ for $r = 5$ and $M_t = 10^{15}$.

Slope $b$ exhibits a distinct maximum which is found for $\mu_{on} \leq 2$. This maximum is accompanied by extreme variance in the number of events occurring within a time interval corresponding to the scaling region [26]. To explain the behavior of slope $b$, we have to postulate that the intermittent g-r process maintains the condition of extreme variance. Table 1 summarizes slope $b$ and the exponent $\mu_{on}$ fulfilling this condition.

The behavior of the maximum value for slope $b$ can be explained by the following considerations: according to Eqs. (A.10) and (A.12), slope $b$ is essentially determined by the quotient $\overline{N_c^2}/\overline{N_c}^2$. The maximum value of this quotient is found for $\mu_{on} = 2$. Substituting Eqs. (A.6) and (30) into Eqs. (A.10) and (A.12), we obtained for

$$r \leq 1: \qquad b \approx 1 - \frac{\log\frac{6}{\pi^2}(\ln M_t + C_E)^2}{\log M_t} \qquad (A.13)$$

and for

$$1 < r \ll \overline{N_c}: \qquad b \approx 1 - \frac{\log\frac{6}{\pi^2}(\ln M_t + C_E)^2}{\log M_t} + \frac{2\log r}{\log M_t}. \qquad (A.14)$$

For $r \leq \overline{N_c}$, the behavior of slope $b$ is described by Eq. (A.11). Herein, the moments of cluster size can be approximated by the following schemes: based on Eqs. (31) and (A.6) and on numerical calculation, the first moment is approximated by

$$1 \leq \mu_{on} \leq 2: \qquad \log \overline{N_c} \approx (2 - \mu_{on}) \log M_t$$
$$2 \leq \mu_{on} \leq 3: \qquad \log \overline{N_c} \approx 0$$

and the second moment by

$$1 \leq \mu_{on} \leq 3: \qquad \log \overline{N_c^2} \approx (3 - \mu_{on}) \log M_t.$$

Applying these approximations, the asymptotic behavior of slope $b$ for large scaling regions as a function of the exponent $\mu_{on}$ is derived by

$$
b \approx
\begin{cases}
1 \leq \mu_{on} \leq 2: & 3 - \mu_{on} \\
2 \leq \mu_{on} \leq 3: & \mu_{on} - 1 + 2\frac{\log r}{\log M_t}.
\end{cases}
$$

The behavior of slope $b$ can therefore be sketched out by two straight lines. The intersection of these two lines provides a rough approximation of the maximum value of slope $b$ by

$$b \approx 1 + \frac{\log r}{\log M_t} \qquad (A.15)$$

which is found at

$$\mu_{on} \approx 2 - \frac{\log r}{\log M_t}. \qquad (A.16)$$